\documentstyle[aps,preprint,times]{revtex}
\draft
\newcommand{\bx}{{\bf x}}
\newcommand{\bu}{{\bf u}}
\newcommand{\ft}{\tilde{f}}
\newcommand{\xivec}{\mbox{\boldmath $\xi$}}
\begin{document}
\title{Discretization of the velocity space in solution of the
Boltzmann equation}
\author{Xiaowen Shan$^{1,2}$ and Xiaoyi He$^{1}$}
\address{$^1$Complex Systems Group, MS-B213, Theoretical Division, Los
Alamos National Laboratory, Los Alamos, New Mexico 87545\\
$^2$PL/GPOF, Air Force Research Laboratory, Hanscom AFB,
Massachusetts, 01731}
\date{September 30, 1997}
\maketitle

\begin{abstract}
We point out an equivalence between the discrete velocity method of
solving the Boltzmann equation, of which the lattice Boltzmann
equation method is a special example, and the approximations to the
Boltzmann equation by a Hermite polynomial expansion.  Discretizing
the Boltzmann equation with a BGK collision term at the velocities
that correspond to the nodes of a Hermite quadrature is shown to be
equivalent to truncating the Hermite expansion of the distribution
function to the corresponding order.  The truncated part of the
distribution has no contribution to the moments of low orders and is
negligible at small Mach numbers.  Higher order approximations to the
Boltzmann equation can be achieved by using more velocities in the
quadrature.
\end{abstract}


The Boltzmann equation is a well accepted mathematical model of a
fluid at the microscopic level.  It describes the evolution of the
single particle distribution function, $f(\bx, \xivec, t)$, in the
phase space $(\bx, \xivec)$, where $\bx$ and $\xivec$ are the position
and velocity vectors respectively.  This description of a fluid is
more fundamental than the Navier-Stokes (NS) equations.  It has a
broader range of application and provides more detailed microscopic
information which is critical for the modeling of the underlying
physics behind complex fluid behavior.  However, direct solution of
the full Boltzmann equation is a formidable task due to the high
dimensions of the distribution and the complexity in the collision
integral.  Among the various techniques developed \cite{Bird94}, the
discrete velocity method was introduced \cite{Broadwell64b} based on
the intuitive assumption that the gas particles can be restricted to
have only a small number of velocities.  The lattice Boltzmann
equation (LBE) method formally falls into this category.

The development of the LBE method for simulation of fluid dynamics was
independent of the continuum Boltzmann equation.  The discrete LBE was
first written to describe the dynamics of the distribution function in
the lattice gas automaton (LGA) \cite{Frisch86,Wolfram86}, in which
the fluid physics is simulated at the microscopic level by ``Boolean''
particles moving with discrete velocities on a regular lattice,
mimicking the motion of the constituent particles of a fluid.  A
Bhatnagar-Gross-Krook (BGK) collision model \cite{Bhatnagar54} was
later adopted in the LBE in place of the complicated collision term
\cite{Chen92a,Qian92a}.  In this lattice Boltzmann BGK model, the
equilibrium distribution is chosen {\em a posteriori} by matching the
coefficients in a small velocity (Mach number) expansion so that the
correct hydrodynamic equations can be derived using the Chapman-Enskog
method.

Recently it has been argued \cite{He97,Abe97} that the LBE method can
be derived from the continuum Boltzmann equation with a BGK collision
model.  In the new derivations, the Maxwellian distribution is Taylor
expanded to second order in the fluid velocity scaled with the sound
speed.  Abe \cite{Abe97} employed a special functional form for the
distribution function so that the macroscopic fluid variables are
completely determined by the values of the distribution function at a
set of discrete velocities.  By noticing that in the Chapman-Enskog
calculation, the functional form of the equilibrium distribution
function in velocity space is only relevant in the calculation of the
low-order moments, and for the Taylor expanded Maxwellian, those
moments can be calculated exactly using a Gaussian quadrature, it is
concluded that the NS equations can be derived from the Boltzmann
equation evaluated on the nodes of the quadrature \cite{He97}.  On
substituting the weights of the corresponding quadrature into the
expansion of the Maxwellian, the coefficients of the LBE equilibrium
distribution function are recovered.  The Boltzmann equation evaluated
at the discrete velocities can then be further discretized in $\bx$
and $t$ in various ways for numerical integration \cite{He97}.  The
LBE models are shown to correspond to solving the discrete Boltzmann
equations with a particular finite difference scheme \cite{Cao97}.

The recovery of the NS equations from the Boltzmann equation by using
a small number of collocation points in velocity space is not
accidental.  Almost half century ago, Grad \cite{Grad49} introduced a
sequence of approximations to the Boltzmann equation by expanding the
distribution function in terms of Hermite polynomials in velocity
space.  The Hermite coefficients are directly related to the
macroscopic fluid variables such as density, velocity, internal
energy, stress and so on.  By keeping Hermite polynomials of up to
third order, Grad obtained a system of equations for thirteen moments
of the distribution function.  This system of equations, known as the
``13 moment'' approximation, was argued to be a better approximation
than the Chapman-Enskog calculation \cite{Grad52,Chapman70}.  By
noticing that the Hermite coefficients for a given function can be
estimated using a Hermite quadrature formula, and that this estimation
is exact when the function satisfies certain conditions, an important
correspondence between the LBE method and the approximation by Hermite
polynomial expansion can be immediately identified.  In this Letter,
we show that by discretizing the Boltzmann-BGK equation at a set of
velocities that correspond to the nodes of a Gauss-Hermite quadrature
in velocity space, we effectively project and solve the Boltzmann
equation in a subspace spanned by the leading Hermite polynomials.
The truncated part of the distribution has no contribution to the
low-order moments that appear explicitly in the conservation
equations.

We start from the following Boltzmann-BGK equation:
\begin{equation}
\frac{\partial f}{\partial t} + \xivec\cdot\nabla f = -\frac 1\tau (f
- f^{(0)}),\label{eq:bgk}
\end{equation}
where, $\tau$ is a relaxation time, $f^{(0)}$ is the Maxwellian
\begin{equation}
f^{(0)} = \rho\left(\frac{m}{2\pi k_BT}\right)^{D/2}e^{-\frac
m{2k_BT}|\xivec - \bu|^2},
\end{equation}
where $D$ is the dimension of the space, $k_B$ is the Boltzmann
constant, and $m$ is the mass of the molecule.  The mass density,
$\rho$, macroscopic fluid velocity, $\bu$, and the temperature, $T$,
are all functions of $\bx$ and $t$.  We introduce the dimensionless
quantity $\theta = Tm_0/T_0m$, where $T_0$ is a characteristic
temperature and $m_0$ is an unit of the molecular mass.  After
rescaling the velocities, $\xivec$ and ${\bf u}$, in units of the
constant $c_0 = \sqrt{k_BT_0/m_0}$, which is the sound speed in a gas
consisting of molecules of mass $m_0$ and at temperature $T_0$, the
Maxwellian takes the following simple form:
\begin{equation}
f^{(0)} = \frac{\rho}{(2\pi\theta)^{D/2}}e^{-\frac 1{2\theta}|\xivec
-\bu|^2}.\label{eq:max}
\end{equation}
The introduction of $m_0$ is for the gas mixtures of components with
different molecular masses.  For a single component system we can
chose $m = m_0$ and have $\theta = T/T_0$.  If the time and length
scales $t_0$ and $L$ are chosen so that $L/t_0 = c_0$, the
dimensionless Boltzmann-BGK equation will have the same form as
Eq.~(\ref{eq:bgk}) with $\tau$ being the dimensionless relaxation
time.  The mass density, $\rho$, the dimensionless fluid velocity,
$\bu$, and the dimensionless internal energy density, $\epsilon =
D\theta/2$, are expressed as the {\em velocity moments} of the form
$\int f\varphi(\xivec)d\xivec$, with $\varphi = 1$, $\xivec$, and
$\xi^2$ respectively:
\begin{equation}
\rho = \int fd\xivec\quad
\rho\bu = \int f\xivec d\xivec\quad
2\rho\epsilon + \rho u^2 = \int f\xi^2d\xivec.
\label{eq:inv}
\end{equation}
In discussions hereafter, the dimensionless variables are used unless
otherwise specified.

Grad introduced the approximations by Hermite expansion in his ``13
moment'' system \cite{Grad49}.  Defining the following weight function
\begin{equation}
\omega(\xivec) = \frac 1{(2\pi)^{D/2}}e^{ - \frac 12\xi^2},\label{eq:w}
\end{equation}
where $\xi^2 = \xivec\cdot\xivec$, as argued by Grad, if
$\omega^{-1/2}f$ is square integrable, {\em i.e.}\ if $f$ approaches
zero faster than $e^{-\xi^2/4}$ as $\xi\rightarrow\infty$, the
following Hermite expansion is valid in the sense of mean convergence:
\begin{equation}
f(\bx, \xivec, t) = \omega(\xivec)\sum_{n = 0}^\infty\frac 1{n!}
a_i^{(n)}(\bx, t){\cal H}_i^{(n)}(\xivec),\label{eq:hexp}
\end{equation}
where ${\cal H}^{(n)}$ is the $n$th order Hermite polynomial
\cite{Grad49b}.  This expansion, also known as the Gram-Charlier
series, was also used in the solution of the Vlasov equation
\cite{Armstrong70}.  Clearly, the Maxwellian in Eq.~(\ref{eq:max}) has
such an expansion if the choice of $T_0$ and $m_0$ ensures that
$\theta < 2$.  On the right-hand-side of Eq.~(\ref{eq:hexp}), both
$a^{(n)}$ and ${\cal H}^{(n)}$ are symmetric tensors of order $n$.
The subscript $i$ is an abbreviation for the multiple indices $\{i_1,
\cdots, i_n\}$, and the products denote contraction on all the $n$
indices.  The Hermite polynomials are a set of complete orthonormal
basis of the Hilbert space with the inner product $\langle f, g\rangle
= \int\omega fgd\xivec$.  They satisfy the orthonormal relation:
\begin{equation}
\int\omega {\cal H}^{(m)}_i{\cal H}^{(n)}_jd\bx = \delta_{mn}
\mbox{\boldmath $\delta$}_{ij},\label{eq:otho}
\end{equation}
where $\mbox{\boldmath $\delta$}_{ij} = 1$ if $i = \{i_1, \cdots,
i_m\}$ is a permutation of $j = \{j_1, \cdots, j_n\}$, and
$\mbox{\boldmath $\delta$}_{ij} = 0$ otherwise.  For any function $f$,
the $n$th Hermite coefficient can be obtained by the following
equation:
\begin{equation}
\int f{\cal H}^{(n)}(\xivec)d\xivec =
\sum_{m=0}^\infty\frac{a^{(m)}}{m!}\int\omega{\cal H}^{(m)}_i
{\cal H}^{(n)}_jd\xivec = a^{(n)},\label{eq:an}
\end{equation}
where the factor $m!$ is the number of terms in $\mbox{\boldmath
$\delta$}_{ij}$.  The function $f$ is completely determined by all of
its Hermite coefficients.

The moments given in Eqs.~(\ref{eq:inv}) are invariants of both the
original Boltzmann collision term and the BGK collision model.  The
hydrodynamic equations are simply the corresponding conservation
equations:
\begin{equation}
\frac\partial{\partial t}\int f\varphi d\xivec + \nabla\cdot\int
f\xivec\varphi d\xivec = 0.
\end{equation}
The lowest order moments have the most significant contribution to the
macroscopic hydrodynamics.  Since the Hermite expansion has the
feature that a velocity moment of a given order is solely determined
by the Hermite coefficients up to that order and are not changed by
the truncation of the higher-order terms, a sequence of approximations
to Eq.~(\ref{eq:bgk}) can be made by seeking the approximate solution
of the following form:
\begin{equation}
\ft(\bx, \xivec, t) = \omega(\xivec)\sum_{n = 0}^N\frac 1{n!}
a_i^{(n)}(\bx, t){\cal H}_i^{(n)}(\xivec).\label{eq:appr}
\end{equation}
The momentum and energy conservation equations explicitly involve
moments of up to the second and the third order respectively.  It is
necessary to require $N\geq 2$ if the momentum equation is to be
obtained and $N\geq 3$ if the energy conservation equation is needed.

By the approximation above, we have assumed that the distribution
function lies entirely in the subspace spanned by Hermite polynomials
up to order $N$.  For higher orders,
\begin{equation}
a^{(n)} \equiv {\bf 0}\quad{\rm if}\quad n > N.\label{eq:trct}
\end{equation}
Although the terms that are truncated do not appear explicitly in the
conservation equations, they affect the fluid variables through their
contributions to the dynamic equations of the lower order moments.  We
will return to the validity of the assumption Eq.~(\ref{eq:trct})
later.

Let $\xivec_i$ and $w_i$, $i = 1, \cdots, d$, be the nodes and weights
of a quadrature of degree $2N$, {\em i.e.}, if $p(\xivec)$ is a
polynomial with a degree not greater than $2N$, we have
\begin{equation}
\int\omega(\xivec)p(\xivec)d\xivec=\sum_{i = 1}^dw_ip(\xivec_i).
\label{eq:cuba}
\end{equation}
Because $\ft{\cal H}^{(n)}/\omega$ is such a polynomial if $n \leq N$,
the Hermite coefficients of $\ft$ can be calculated using the values
of $\ft$ at the nodes $\xivec_i$ as the following:
\begin{equation}
a^{(n)} = \int\omega\frac\ft\omega{\cal H}^{(n)}d\xivec =
\sum_{i = 1}^d\frac{w_if_i{\cal H}^{(n)}(\xivec_i)}{\omega(\xivec_i)},
\end{equation}
where, $f_i = \ft(\bx, \xivec_i, t)$.  The knowledge of $f_i$ as
functions of position and time is equivalent to that of the truncated
distribution function itself and therefore, that of the fluid
variables calculated from the truncated distribution.  These variables
are:
\begin{eqnarray}
\rho = \sum_{i = 1}^d\frac{w_if_i}{\omega(\xivec_i)},\quad &&
\rho\bu = \sum_{i = 1}^d\frac{w_if_i\xivec_i}{\omega(\xivec_i)},\nonumber\\
2\rho\epsilon + \rho u^2 &=& \sum_{i = 1}^d
\frac{w_if_i\xi_i^2}{\omega(\xivec_i)}.\label{eq:macro}
\end{eqnarray}
By defining the auxiliary variables $g_i = w_if_i/\omega(\xivec_i)$,
Eqs.~(\ref{eq:macro}) can be put into a more efficient form for
computation:
\begin{equation}
\rho = \sum_{i = 1}^dg_i, \quad
\rho\bu = \sum_{i = 1}^dg_i\xivec_i, \quad
2\rho\epsilon + \rho u^2 = \sum_{i = 1}^dg_i\xi_i^2.\label{eq:macro2}
\end{equation}
This has the same form as how the fluid variables are calculated in
the LBE method, where the distribution function is defined from the
beginning as the populations of particles moving at discrete
velocities.

We now turn to the discussion of the equations that the functions
$f_i$ satisfy.  By directly evaluating Eq.~(\ref{eq:bgk}) at the nodes
$\xivec_i$, we have:
\begin{equation}
\frac{\partial f_i}{\partial t} + \xivec_i\cdot\nabla f_i =
-\frac 1\tau[f_i - f^{(0)}(\xivec_i)].
\label{eq:dbe}
\end{equation}
Because $f^{(0)}$ has non-zero Hermite coefficients at all orders, on
substituting $f^{(0)}(\xivec_i)$ into the right hand side of
Eqs.~(\ref{eq:macro}), the equalities hold only approximately.
Clearly the mass density, momentum and total energy will not be
exactly conserved by the collision term on the right hand side of
Eq.~(\ref{eq:dbe}).  For the conservation laws to hold exactly,
$f^{(0)}$ has to be projected into the subspace in which $\ft$ lies.
Namely, in Eq.~(\ref{eq:dbe}), the Maxwellian has to be replaced by
its $N$th order Hermite expansion.  Denoting the values of the
auxiliary variables $g_i$ corresponding to the truncated Maxwellian by
$\tilde{g}^{(N)}_i$, Eq.~(\ref{eq:dbe}) can be written as
\begin{equation}
\frac{\partial g_i}{\partial t} + \xivec_i\cdot\nabla g_i = -\frac
1\tau(g_i - \tilde{g}^{(N)}_i),\quad i = 1, \cdots, d.\label{eq:lbe}
\end{equation}
It is to be noted that the positivity of the distribution function is
lost in this truncation.

The first a few Hermite coefficients of the Maxwellian can be obtained
using Eq.~(\ref{eq:an}).  They are: $a^{(0)} = \rho$, $a^{(1)}_i =
\rho u_i$, $a^{(2)}_{ij} = \rho u_iu_j + (\theta - 1)\rho\delta_{ij}$,
and $a^{(3)}_{ijk} = \rho u_iu_ju_k + (\theta - 1)\rho(u_i\delta_{jk}
+ u_j\delta_{ik} + u_k\delta_{ij})$.  Using the explicit form of the
Hermite polynomials, at the second and the third orders, we have
\begin{eqnarray}
\tilde{g}^{(2)}_i &=& w_i\rho\left[1 + \gamma_i + \xivec_i\cdot\bu +
\frac 12(\xivec_i\cdot\bu)^2 - \frac{u^2}2\right],\label{eq:g01}\\
\tilde{g}^{(3)}_i&=& w_i\rho\left[1 + \gamma_i + (1 + \zeta_i)
\xivec_i\cdot\bu + \frac
12(\xivec_i\cdot\bu)^2 - \frac{u^2}2\right.\nonumber\\
&&\left.+\frac 16(\xivec_i\cdot\bu)^3 -\frac
12(\xivec_i\cdot\bu)u^2\right],\label{eq:g00}
\end{eqnarray}
where $\gamma_i = (\xi_i^2 - D)(\theta - 1)/2$, $\zeta_i = (\xi_i^2 -
D - 2)(\theta - 1)/2$.  Eqs.~(\ref{eq:lbe})--(\ref{eq:g00}) are the
projection of the Boltzmann-BGK equation in the subspace spanned by
the leading Hermite polynomials.  They are in the configuration space
$(t, \bx)$ and have a linear differential operator on the left-hand
side.  The fluid variables defined by Eqs.~(\ref{eq:macro2}) obeys the
NS hydrodynamics.  As previously shown \cite{He97,Abe97}, the LBE
equilibrium distributions of Refs.~\cite{Chen92a,Qian92a} are obtained
when the proper nodes and weights are substituted into
Eq.~(\ref{eq:g01}), and the lattice Boltzmann equations are particular
finite difference discretizations of Eq.~(\ref{eq:lbe}).

It can be easily verified that moments of up to second and third
orders calculated from Eqs.~(\ref{eq:g01}) and (\ref{eq:g00})
respectively are those of the Maxwellian.  In particular, the tensor,
$p_{ij}\equiv\int f^{(0)}\xi_i\xi_j d\xivec = \rho u_iu_j +
\rho\theta\delta_{ij}$, survives the truncation.  The hydrostatic
pressure is given by the dimensionless equation of state, $p =
\rho\theta$, which translates to $p = \frac{\rho}{m}k_BT$ in
laboratory units, and the sound speed is $\sqrt{\theta}$ as expected.
When measured in the magnitude of one of the nodes of the quadrature,
{\em e.g.}\ $\xi_i$, as in the LBE models, the sound speed is
$\sqrt{\theta}/\xi_i$.  In a single component isothermal system,
$\theta$ becomes a free parameter which can be used to adjust the
nominal sound speed with respect to the nodes of the quadrature.  When
$\theta = 1$, the Maxwellian has a very simple expansion:
\begin{equation}
f^{(0)} = \omega\rho\sum_{n=0}^\infty\frac 1{n!}\bu^{(n)}{\cal
H}^{(n)}.\label{eq:f0}
\end{equation}
and the truncated part of the distribution function is proportional to
the power of the Mach number.  Eqs.~(\ref{eq:g01})--(\ref{eq:g00}) are
also simplified since $\gamma_i = \zeta_i = 0$.  For a multiple
component system, a different $\theta\sim 1/m$ has to be chosen for
each component if all the components are at thermal equilibrium.  This
requirement was found necessary to obtain a correct equation of state
in a previous multiple component LBE model \cite{Shan96}.

It should be noted that the truncation made in Eq.~(\ref{eq:trct}) is
similar to, but not exactly the same as the third order approximation
in the Grad ``13 moment'' system.  In the latter the distribution
function is expanded in velocity space around the local fluid velocity
before it is truncated.  This is certainly a better approximation than
expanding around the absolute equilibrium, and corresponding
computational methods can be developed.  However, it is not possible
to use such an expansion in the method discussed above because it
would yield a set of nodes that depend on the local velocity.

The difference between the two expansions can be estimated by
expanding the following approximated distribution function of the ``13
moment'' system \cite{Grad49} around the absolute equilibrium:
\begin{equation}
f^{(0)}\sum_{i = 0}^3\frac{b^{(i)}}{i!}{\cal H}^{(i)}(\xivec - \bu).
\end{equation}
where, $b^{(0)} = 1$, $b^{(1)} = 0$, and $b^{(2)}_{ii} = 0$.  The
Hermite coefficients in the expansion around the absolute equilibrium,
$a^{(n)}$, can be calculated as the following:
\begin{eqnarray}
a^{(n)} &=& \int f^{(0)}\sum_{i = 0}^3\frac{b^{(i)}}{i!}{\cal H}^{(i)}
(\xivec - \bu){\cal H}^{(n)}(\xivec)d\xivec\\
&=&\rho\sum_{i = 0}^3\frac{b^{(i)}}{i!}\int\omega(\xivec)
{\cal H}^{(i)}(\xivec){\cal H}^{(n)}(\xivec+\bu)d\xivec
\end{eqnarray}
Noticing that ${\cal H}^{(n)}(\xivec + \bu) = \sum_{i = 0}^n \bu^{(n -
i)}{\cal H}^{(i)}(\xivec)$, we find that $a^{(0)} = \rho$, $a^{(1)} =
\rho\bu$, $a^{(2)} = \rho(\bu^{(2)} + b^{(2)})$, and for $n \geq 3$,
\begin{equation}
a^{(n)} = \rho(\bu^{(n)} + \bu^{(n - 2)}b^{(2)} + \bu^{(n - 3)}b^{(3)}).
\end{equation}
With non-zero Hermite coefficients at all orders, the distribution
function in the ``13 moment'' system does not meet the assumption in
Eq.~(\ref{eq:trct}).  Since these coefficients are proportional to the
power of the Mach number, Eqs.~(\ref{eq:g00}) approximate the ``13
moment'' system only at the small Mach number limit.  This
approximation can be improved by using a quadrature formula with a
higher degree.

Except for one dimension, finding the quadrature formula with minimum
number of nodes for given geometry, weight function, and degree of
accuracy is generally an unsolved problem.  As the minimum
requirement, a formula of 4th degree is needed for isothermal systems
and 6th degree for thermodynamic systems.  For the weight function in
Eq.~(\ref{eq:w}), quadrature formulae of different degrees are listed
in Ref.~\cite{Stroud71}.  Some of those are believed to be minimum
without proof.  In two dimensions, the minimum formula seems to be the
4th degree, 6-point formula (origin and the vertices of a pentagon)
for isothermal models and the 7th degree, 12-point formula for
thermodynamics models.  In three dimensions, the minimum formulae are
those of 5th degree, 13-point (origin and the vertices of a regular
icosahedron) and 7th degree, 27-point for isothermal and thermodynamic
systems respectively.  The nodes of these formulae usually do not
coincide with that of a regular lattice.  Eqs.~(\ref{eq:lbe}) have to
be solved using schemes such as the finite difference method
\cite{He97,Cao97}.

The methodology of the discrete Boltzmann equation can be summarized
as the following.  The discretization of the continuum distribution
function into values at the nodes of a quadrature formula is
equivalent to the truncation of the high-order terms in the Hermite
spectral space.  The information that is lost in this procedure is
represented by the high-order Hermite polynomials which do not
explicitly appear in the conservation equations.  This error is
negligible at small Mach numbers and can always be made smaller by
using a quadrature of a higher degree.  With such a discretization,
the Boltzmann equation becomes a homogeneous set of linear equations
in the configuration space.  Comparing with the non-linear NS
equations, these equations are easier to solve, have a broader range
of application, and more importantly, allow the underlying fluid
physics to be simulated directly at the cost of a macroscopic
simulation.  In addition, higher order approximations to the Boltzmann
equation can be easily achieved by adding more points to the system.

Some of the limitations that LBE methods inherited from the Boolean
LGA models can be removed with the present formulation.  The
equilibrium distribution is now obtained through a systematic
orthogonal expansion of the Maxwellian, eliminating the tedious
parameter-matching procedure which usually produces results that are
not unique and contain erroneous terms at higher orders.  The
inflexible lattice structure and time stepping scheme of the LBE
method are inconvenient for practical applications and often result in
poor stability.  By realizing that the LBE models are merely simple
and rather primitive finite difference representations of the discrete
Boltzmann-BGK equation, we can employ more sophisticated numerical
techniques in solving these equations with better efficiency,
stability and flexibility.  We defer the detailed discussion and
numerical examples to a future publication.

The authors thank Dr.~Gary Doolen and Dr.~Nicos Martys for helpful
discussions.


\end{document}